# Three-dimensional characterization of the steel-concrete interface by FIB-SEM nanotomography

**Authors:** Nicolas RUFFRAY[1], Ueli M. ANGST[1]*, Thilo SCHMID[1], Zhidong ZHANG[1], and O. Burkan ISGOR[2]

[1] Institute for Building Materials, ETH Zurich, Switzerland
[2] School of Civil and Construction Engineering, Oregon State University, USA

* U. Angst (corresponding author)
ETH Zurich, Institute for Building Materials, Laura-Hezner-Weg 7, CH-8093 Zurich, Switzerland
Tel: +41 44 633 40 24, e-mail: uangst@ethz.ch

**Abstract**
While it is widely accepted that the steel-concrete interface (SCI) plays an important role in governing the long-term durability of reinforced concrete structures, understanding about the primary features of the SCI that influence corrosion degradation mechanisms has remained elusive. This lack of knowledge can be attributed, on the one hand, to the complex heterogeneous nature of the SCI, and, on the other hand, the absence of experimental techniques suitable for studying the relevant features of the SCI. Here, we use focused ion beam – scanning electron microscopy (FIB-SEM) nanotomography to obtain high-resolution 3D tomograms of the steel-concrete interfacial zone. Five tomograms, spanning volumes ranging from 8,000 to 200,000 µm³, were acquired for situations representative of both non-corroded and corroded SCIs. The achieved voxel size falls within the range of 30–50 nm, thus providing a resolution clearly surpassing the capabilities of computed X-ray tomography. This resolution enables the 3D characterization of the microstructure at the capillary scale, which is the scale at which relevant corrosion and related mass transport processes occur. Thus, FIB-SEM nanotomography is capable of yielding datasets of the SCI that serve as basis for the generation of digital twins of the interfacial microstructure, thereby enabling future studies about durability and corrosion of reinforced concrete at the pore scale.

# 1. Introduction

It is widely recognized that the steel-concrete interface (SCI) plays a crucial role in the long-term performance and durability of reinforced concrete structures [1-3]. Various microstructural features related to the SCI, such as phase assemblage and pore solution chemistry [4-9], pore structure [10, 11], macroscopic defects [12-18] and interfacial moisture distribution [19-23] have been identified to strongly affect corrosion of the embedded steel. However, despite extensive research spanning several decades, the primary characteristics of the SCI that govern the steel corrosion mechanisms fundamentally are still largely unclear [2].

This lack of knowledge can be attributed to the complex and heterogeneous nature of the SCI [1]. Additionally, another important reason contributing to the limited understanding and, to some extent, the contradictory findings in the literature [2] is the absence of well-established experimental techniques suitable for studying the SCI and, especially, techniques capable of revealing the features of relevance for the degradation mechanisms affecting the long-term durability of reinforced concrete. In a recent extensive literature review [24] over 20 experimental methods were critically assessed with respect to their aptness for the study of relevant features of the SCI. The review revealed a severe lack of established techniques suitable for characterising the SCI at a representative scale and with sufficiently high resolution. A particularly challenging aspect identified in [24] included the fact that the microstructure of the SCI is three-dimensional and that 2D imaging techniques, e.g. ex-situ microscopic techniques such as optical or scanning electron microscopy (SEM), are generally not capable of providing such 3D information. This limitation of viewing features only in 2D sectional planes may lead to misinterpretation, for instance, with respect to interconnectivity of pores or the size of features such as quasi-spherical voids. Three-dimensional characterisation techniques exist, among which computed X-ray tomography currently seems the most widely used one [17, 25-27], followed by a few applications of neutron tomography or combined neutron and X-ray tomography [18, 28]. While these techniques can reveal interesting 3D features of the microstructure at the SCI, their limited spatial resolution poses a challenge in resolving the SCI at a scale below approx. 10 micrometers [24]. This limitation applies in particular because of constraints about the size of samples that can be retrieved from the steel-concrete interfacial zone. It is extremely challenging, if not virtually impossible, to obtain specimens of the interfacial region that would be small enough to allow for submicron X-ray tomography, without risking significant damage of the features of interest during sample preparation. Another weakness, particularly of X-ray computed tomography, is its sensitivity to imaging artefacts arising from the pronounced differences in X-ray attenuation coefficients of the steel and cementitious matrix, that may give rise to X-ray beam hardening effects of polychromatic X-ray beams [29-32], which in turn may impair the interpretation and analysis of the obtained images.

Another imaging technique that is capable of yielding three-dimensional information and at much better spatial resolution than X-ray and neutron tomography is *focused ion beam – scanning electron microscopy tomography*, generally abbreviated as FIB-SEM tomography. The approach essentially consists in serial milling of sections by means of a FIB to generate stacks of sections with nanometer precision, imaging of each section by SEM, and using these SEM micrographs to create a 3D image with a spatial resolution in the nanometer scale. Another advantage is that the FIB milling typically introduces fewer artifacts than mechanical polishing methods [24]. FIB-SEM tomography has been employed to study cementitious materials [33-35], and preliminary results of an exploratory study using FIB-SEM tomography to characterise the SCI were reported in [24].

This work aims to further explore the potential of FIB-SEM nanotomography as a technique to characterise the microstructure of the SCI. We consider FIB-SEM nanotomography a particularly powerful and promising technique since it can provide 3D information about the microstructure of the interfacial cementitious matrix, including the distribution of different hydrated and unhydrated cementitious phases, aggregates, and the pore structure in the scale where fundamental durability-relevant processes occur. These processes include, for instance, capillary condensation and water transport, ion transport, crystallization and precipitation of corrosion products, and poromechanical processes.

In this paper, we illustrate the potential of FIB-SEM tomography using two different types of specimens: An "uncorroded SCI" that may be representative for situations prior to initiation of corrosion, and a "corroded SCI" that may represent the situation after corrosion initiation, where the characterization of other features, including the presence of rust in the matrix, are of interest in developing fundamental understanding about SCI.

## 2. Materials and Methods

### 2.1. Specimens

Two types of specimens were studied in this work (Figure 1). These specimens were not intended to be compared with each other directly, rather they were prepared to specifically study an "uncorroded SCI" and a "corroded SCI" independently.

The first type of specimen, designated "NCI" for "non-corroded interface", contained a stainless-steel tube (diameter 3.0 mm, wall thickness 0.25 mm, DIN 1.4301) embedded in mortar matrix produced from ordinary Portland cement (CEM I 42.5N), with a water/cement ratio of 0.5 and a sand/cement ratio of 2. The sand was a pure limestone sand with maximum particle size of 2 mm. After mixing, the mortar was poured in a cubic mould of side length 40 mm, in which the stainless-steel tube was vertically fixed at the centre. Subsequently, the moulds were vibrated for 20 s on a vibrating table. After 1-d curing, they were demoulded and curing continued at 95% relative humidity for two years. The purpose of using stainless-steel as reinforcement in this case, and a mature Portland cement system, was to avoid corrosion at the SCI, and thus to allow for the study of an "non-corroded SCI". More information about the specimen production can be found in [29].

The second specimen type, designated "CI" for "corroded interface", describes concrete with an embedded carbon steel reinforcing steel bar (diameter 6 mm, B500B). The concrete was produced from a slag cement (CEM III/B) with water/binder ratio = 0.5 and sand/binder ratio = 3 by weight. The maximum aggregate size was 8 mm. A specimens of dimensions 15x15x4 $cm^3$ was cast with three reinforcing steel bars embedded at cover depth 15 mm from both sides. The specimen was demoulded after one day and cured at 95% relative humidity (RH) for 3 months. Subsequently, it was subjected accelerated carbonation at 65–75% RH and 50–100% $CO_2$ concentration until the entire specimen was fully carbonated (which was validated by companion specimens that were periodically split and sprayed with phenolphthalein). Once carbonated, the specimen was allowed to corrode in the following conditions: at 95% RH as well as under cyclic wetting/drying exposure (up to 6 cycles). For the latter, the specimen was exposed to demineralized water from one side (max. 7 days), followed by exposure to 30–40 % relative humidity. The purpose of this procedure – namely using carbon steel, slag cement, accelerated carbonation, and exposure to different moisture conditions – was to generate a situation where the steel is corroded under carbonation conditions and with a variable moisture exposure history, thus leading to a certain amount of corrosion products distributed in the cementitious matrix surrounding the steel ("corroded SCI").

All the experiments described above were performed at indoor laboratory temperature ranging between 20 and 23 °C. Table 1 provides a summary of the properties of two specimens studied in this research.

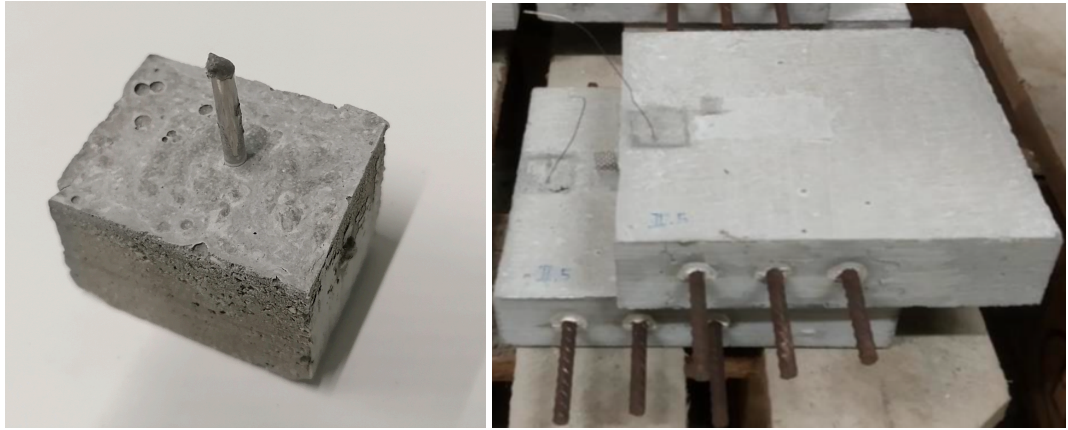

*Figure 1. Photographs of the two different types of specimens used. a) stainless steel tube embedded in Portland cement mortar (non-corroded interface); b) carbon steel reinforcing bars embedded in carbonated slag-cement concrete (corroded interface). For dimensions, it is referred to Table 1.*

*Table 1. Material specifications of the two types of specimens used*

| Specimen type | NCI | CI |
|---|---|---|
| | Non-corroded SCI | Corroded SCI |
| Binder type | OPC, CEM I 41.5 N | Slag cement, CEM III/B |
| w/b | 0.5 | 0.5 |
| Aggregate/binder ratio | 2 | 3 |
| Aggregate max. diameter | 2 mm | 8 mm |
| Steel type | Austenitic stainless steel (DIN 1.4301) | Carbon steel |
| Steel geometry | Tube, diameter 3 mm, wall thickness 0.25 mm | Ribbed reinforcing steel, diameter 6 mm |
| Remarks | Well cured and high degree of hydration of the cementitious matrix | Subjected to accelerated carbonation and variable moisture exposure history |

### 2.2. Specimen preparation for microscopy

To prepare the specimens for microscopy, the specimens were first cut, using a water-cooled diamond circular blade, perpendicular to the reinforcement axis, leaving a thickness of 10 to 15 mm to match the height limitations of the microscopes used in this study. The edges of the cuts were then trimmed to fit inside 1 inch in diameter silicone moulds used for epoxy vacuum impregnation. After being allowed to dry for at least 48 hours in a desiccator under vacuum containing desiccant beads, the specimens were epoxy vacuum impregnated using *Epo-Tek 301* epoxy resin in a *CitoVac* vacuum impregnation unit from *Struers*. After at least 24 hours of curing at ambient temperature the samples were progressively polished using different grades of sandpaper (500 and 1000 grit) and diamond paste (9, 3 and 1 $\mu m$). The cleaning of the specimens in between each grade of polishing was performed in an ethanol ultrasound bath. Cutting, vacuum impregnation and polishing of the specimens were all performed to assure that the axis of the reinforcement was perpendicular to the face that would be later observed in the microscopes. After polishing, the specimens were stored to dry in a desiccator for another 24 hours prior to mounting. The specimens were then mounted on 1-

Pre-print, October 2023　　　　　　　　　　　　　　　　　　　　　　　　　　　　　　　　　　　4

in aluminium stubs using *Conductive Carbon Cement* and *Silver Paint* to assure a proper electrical conductivity between the polished face and the mounting stub. Finally, after allowing the mounting materials to dry for 24 hours, the specimens were carbon coated using a *Bal-Tec CED 030* carbon evaporator. Carbon coating was preferred over metallic coatings to assure a better Back-Scatter Electron (BSE) imaging of the polished surface.

### 2.3. FIB-SEM tomography

Two different SEM with similar capabilities and confirmed output performance were used. The Helios 600i from *FEI* (referred to as 600i in the rest of this article) was used to obtain the tomograms of the NCI specimen, while the Helios 5 UX from *Thermo Fisher* (referred to as 5 UX) was used on CI type of specimens. Both SEM were equipped with a secondary Gallium Focused Ion Beam (FIB) positioned at an angle of 52° relative to the electron beam (Figure 2b). For both microscopes the SEM eucentric and SEM-FIB coincidence points were located at a working distance of 4.1 mm.

Figure 2 presents a schematised construction of the microscopes and the locations of different BSE and Secondary Electron (SE) detectors used during this study. Only the 5 UX was equipped with an In-column Detector (ICD) while other detectors are common to both microscopes. Note that while the Concentric Back-Scatter detector (CBS) produces the most qualitative BSE images, due to its retractable nature, it cannot be used when the sample is tilted at 52° as it would be in the way of the microscope stage.

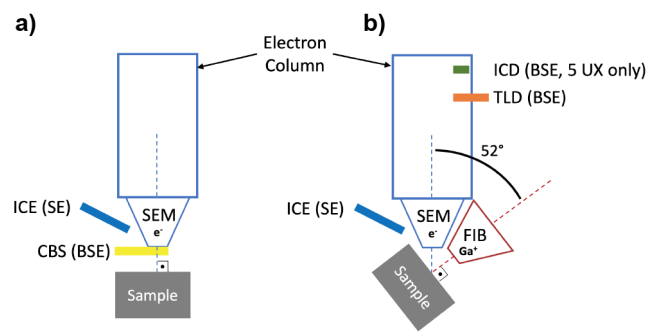

*Figure 2. Setup of microscopes with different detectors: Ion Conversion and Electron detector (ICE), Concentric Back-Scatter detector (CBS), Through-Lens Detector (TLD), and In-Column Detector (ICD, only equipped on the 5 UX). a) Detectors used at 0° tilt in a scanning electron microscope (SEM) with CBS. b) Detectors used at 52° tilt in a setup combined SEM with a focused ion beam (FIB).*

Figure 3 shows a schematic representation the FIB-SEM tomography preparation and process [36]. First, a Region of Interest (ROI) was chosen for tomography by observing the specimen's polished surface, in our case the steel-concrete interface. The selected area was then covered by a platinum or tungsten protective layer using the gas injection system of the microscope in combination with the FIB. The role of this layer was to protect the volume of interest from being damaged by FIB during the tomography preparation and the tomography itself as a FIB image is taken at every tomography iteration for alignment of the ion beam. A trench was opened in front of the volume of interest with the FIB exposing the imaging plane to the electron beam. Note that optional side trenches could also be opened to allow a better evacuation of the milled away matter and minimize the re-depot growth. A FIB and optional SEM fiducials were engraved to allow beams alignment during the tomography process. Finally, the imaging plane was polished at low FIB currents to obtain a smooth surface for imaging.

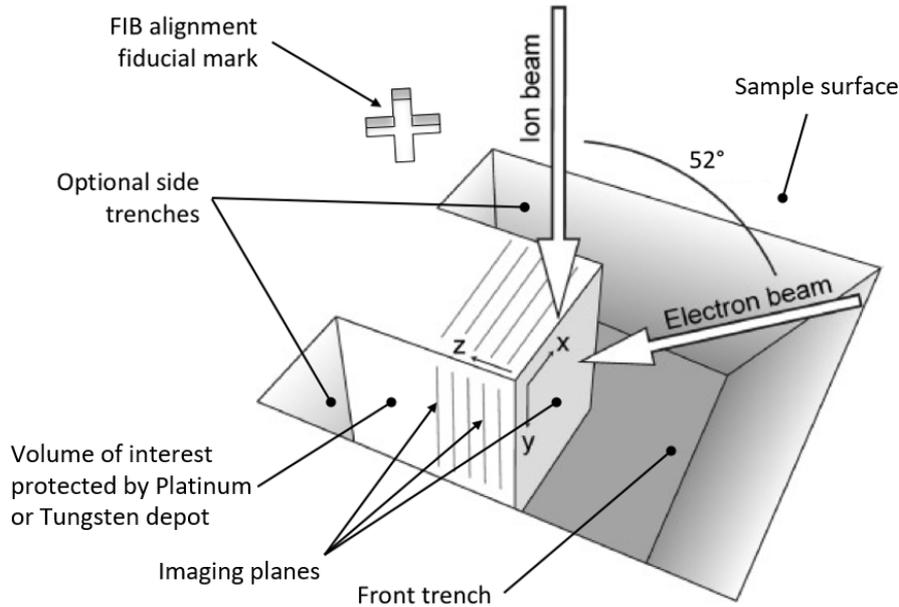

*Figure 3. Schematic illustration of the FIB-SEM tomography process (adapted from [36]).*

The tomography imaging process was then started and consisted in an automatic iteration of milling away a slice of the volume of interest and taking an image of the exposed plane. By repeating this process over all the volume of interest, a stack of images was obtained which can be later used for three-dimensional segmentation and characterization of the microstructure.

### 2.3.1. FIB Milling

All milling operations were performed with the Gallium FIB at a voltage of 30 kV. The opening of the trenches was done with a current of 65 nA. Polishing of the imaging plane was realised with currents of 9.1 and 2.5 nA allowing to obtain an acceptable smoothness for imaging. Tomographies were performed with a milling current of 9.1 nA and a slice thickness of either 30 or 50 nm.

### 2.3.2. SEM Imaging

For general observations and navigation of the specimens a high-performance Ion Conversion and Electron (ICE) detector was used at voltages between 3 and 4 kV and currents between 80 pA to 3.2 nA. This SE detector can be used at any tilt angle of the specimen without noticeable loss of image quality.

At a 0° tilt angle of the specimen a Concentric Back-Scatter (CBS) detector was used to acquire BSE images to determine the ROI where the tomographies were performed. The CBS detector was operated at voltages of 5 to 10 kV and currents of 1.6 to 3.2 nA. It is important to note that while the CBS detector is able to provide the best quality of BSE images among all available detectors, it cannot be used when the specimen is tilted at 52° for the tomography. This is because the CBS detector is a retractable one and that in its deployed position the tiled specimen would collide with it as can be observed in Figure 2.

To acquire BSE tomography images in the tilted position of the specimen, either the Through-Lens (TLD) or the In-Column (ICD) detectors were used. TLD image acquisition was released at a voltage of 2 kV, a current of 0.34 nA and a dwell time of 3 us with a line integration of 8. For ICD image acquisition a voltage of 4 kV, a current of 3.2 nA and a dwell time of 4 us with a line integration of 2 were used.

These parameters were empirically obtained to achieve an acceptable compromise between image quality and acquisition time. The horizontal field width (magnification) and the pixel resolution of the images were chosen to obtain a pixel size of either 30 or 50 nm allowing together with the tomography slice thickness to achieve isometric voxels of the same size. During the tomography images acquisitions, both tilt correction and dynamic focus were used to compensate the 38° angle between the electron beam and the imaging plane normal.

To increase the imaging quality of in-column located detectors such as TLD and ICD, both microscopes used in this study possess a proprietary technology called Immersion Mode. When active, this mode allows the detection of a greater number of electrons by using an electro-magnetic field applied at the tip of the electron column (Figure 4). It is important to note here that during the tomography process, the TLD detector requires the usage of immersion mode to produce exploitable images. While being very beneficial for the quality of the produced images, immersion mode presents the major disadvantage of not being usable if the specimen contains large amount of ferro-magnetic metal as discussed in section 3.3.2.

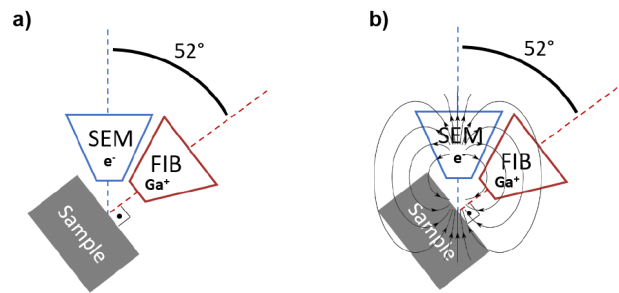

*Figure 4. Detection Modes. a) Normal Mode; b) Immersion Mode. In Immersion Mode, an electro-magnetic field is applied at the tip of the electron column to increase the signal (eq. number of electrons) of the detectors located in the SEM column (TLD and ICD).*

# 3. Results and Discussion

The objective of this section is to present the tomograms obtained during this study as well as the various challenges faced during this process. The processes of image treatment, three-dimensional segmentation and analysis of the obtained data to generate digital twins of the SCI will be presented and discussed in a separate publication.

## 3.1. Tomograms

Five tomograms were generated during this study – four for the NCI specimen, and one for the CI specimen. The key data for each tomograph are presented in Table 2.

*Table 2. Key data of the five obtained tomograms.*

| Specimen | NCI-1 | NCI-2 | NCI-3 | NCI-4 | CI |
|---|---|---|---|---|---|
| Voxel size [nm] | 30 | 50 | 50 | 30 | 30 |
| Microscope | 600i | 600i | 600i | 600i | 5 UX |
| Detector | TLD | TLD | TLD | TLD | ICD |
| Detection Mode | Immersion | Immersion | Immersion | Immersion | Normal |
| Size X [µm] | 28.5 | 53.8 | 45.0 | 48.0 | 55.5 |
| Size Y [µm] | 19.0 | 32.8 | 30.0 | 32.4 | 71.9 |
| Size Z [µm] | 15.4 | 13.1 | 29.5 | 35.9 | 53.8 |
| Tot. Vol. [$\mu m^3$] | 8'339 | 23'029 | 39'825 | 55'793 | 214'686 |

Figure 5 shows examples of a three-dimensional reconstruction that can be obtained from the tomograms. Note that the tomograms shown in Figure 5c,d,e are composed of blocks exhibiting different grey values. These variations in grey values stem from the fact that these "large" tomograms were stitched together from several sub-tomograms that were acquired on the microscopes throughout several runs. This is due to limited availability of the imaging equipment, which was generally limited to half days and weekends, combined with the long measuring times. As a result, the volumes of the size of interest in this work could only be acquired over several interrupted runs. As an example, for the tomogram CI, 8 FIB-SEM sessions were needed with a total of approximatively 116 hours of acquisition time, and additional 20 to 25 hours of sample preparation on the microscope prior to the FIB-SEM tomography measurements. At the beginning of each run, attempts were always made to adjust the brightness and contrast such that the first image of the new run looked as similar as possible to the last image of the previous run. However, this obviously remained to some extent subjective. Another influencing factor causing the differences in grey values during acquisition were drifts in brightness and contrast related to the devices, which, in this study, was found to be particularly pronounced on the 5UX instrument, but also occurring on the 600i device, although to a smaller extent. For these reasons, it is challenging to ensure constant image brightness and contrast during acquisition. Thus, these features have to be corrected for during subsequent image treatment.

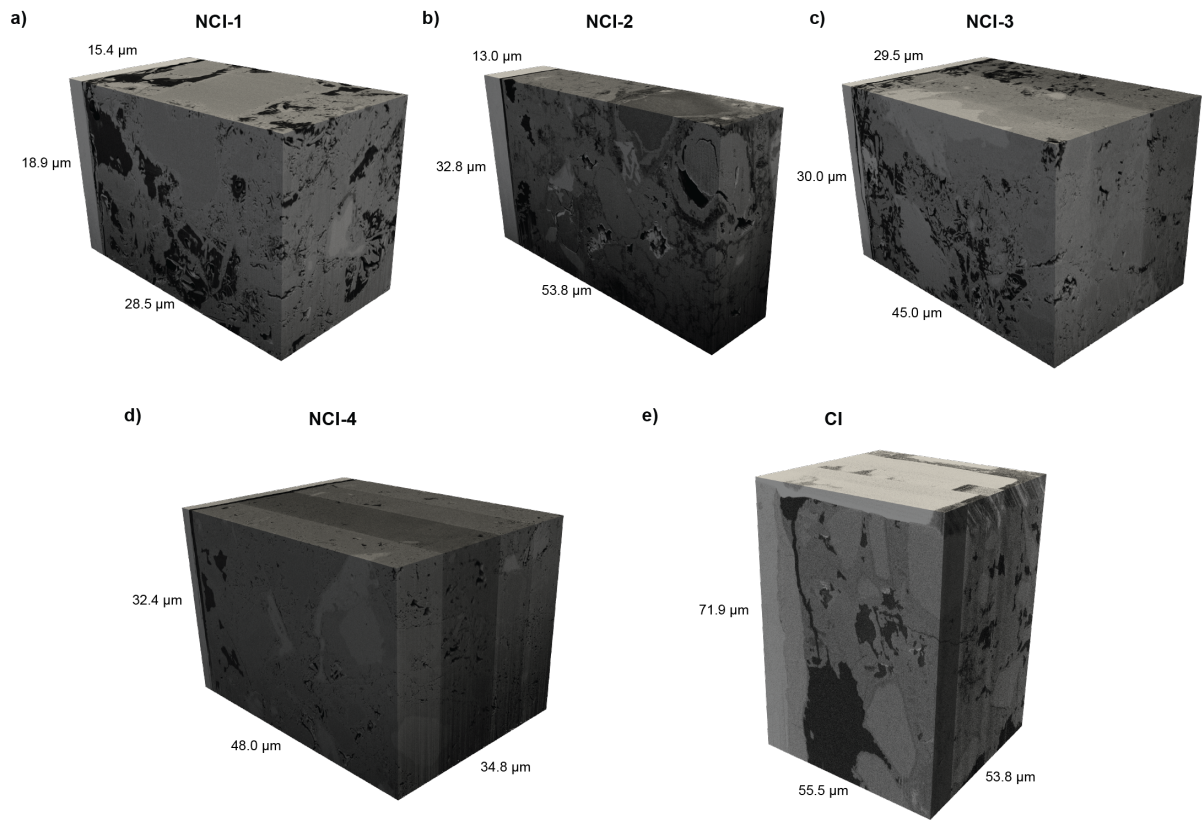

*Figure 5. Examples of three-dimensional renderings of the tomograms of the four samples representing a non-corroded interface (NCI) (a,b,c,d) and the sample representing the corroded interface (CI) (e).*

A typical tomography slice (SEM imaged section) can be observed in Figure 6. On this image of NCI-3 obtained by TLD detector in immersion mode, the bright vertical strip on the left is the stainless-steel tube while the rest of the image shows the adjacent mortar microstructure. Cement hydrates appears as darker levels of grey while unhydrated cement exhibits lighter ones. Epoxy resin filled porosity appear black.

From this image, one can observe the complex structure of porosity present at the steel-mortar interface as well as in the nearby transition zone. The choice of the voxel size is important to be able to observe and later segment small features such as capillary pores. Indeed, derived from to the Nyquist sampling theorem [37, 38], it is considered that a feature must be covered by at least 2.3 adjacent pixels to be segmented as such. Therefore, with this ratio assumed to be 3, the obtained tomograms present a feature resolution of 150 and 90 nm for 50 and 30 nm voxel sizes, respectively. Upon 3D reconstruction and segmentation, the tomography data can provide detailed and interesting information on parameters such as pore size distribution, geometry, interconnectivity, and tortuosity.

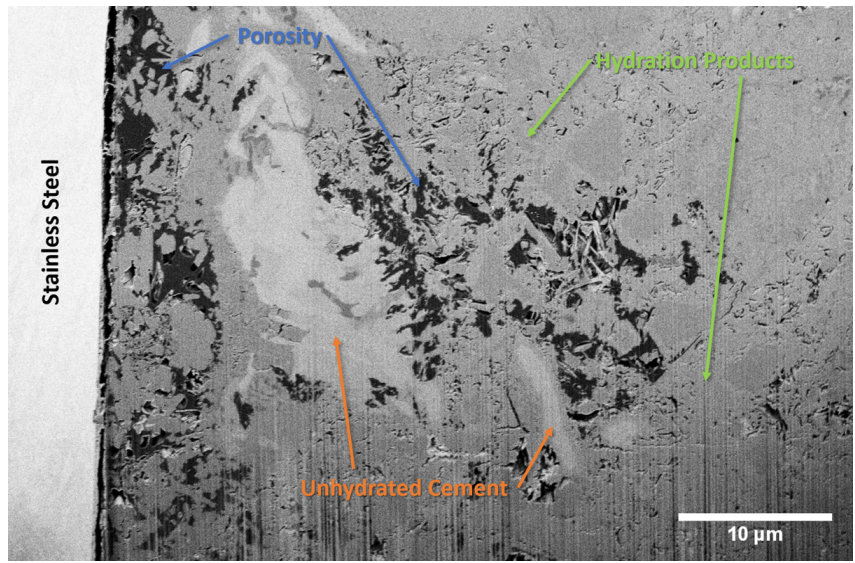

*Figure 6. Tomogram slice (SEM imaged section) of specimen NCI-3. Obtained with the Helios 600i (TLD, Immersion Mode, 2 kV, 0.34 nA, 52° tilt). The curtaining effect observed at the bottom of the image comes from the relatively important depth of the tomogram and the presence porosity that was not filled with epoxy resin, therefore creating a hardness inhomogeneity resulting in this artefact.*

Figure 7 shows a tomogram slice of specimen CI, obtained by ICD detector in normal mode. Like Figure 6 the bright vertical area on the left of the image represents the edge of the reinforcement. Adjacent to this zone, one can observe a slightly darker one which in this case corresponds to corrosion products.

By comparing Figure 6 and Figure 7, it is also possible to notice some differences linked to the use of different detectors. TLD in immersion mode tends to produce sharp images with a good contrast range while ICD in normal mode shows a certain amount of noise. Nevertheless, the higher level of noise of the ICD can be dealt without any loss of image quality and resolution during the image processing prior to segmentation.

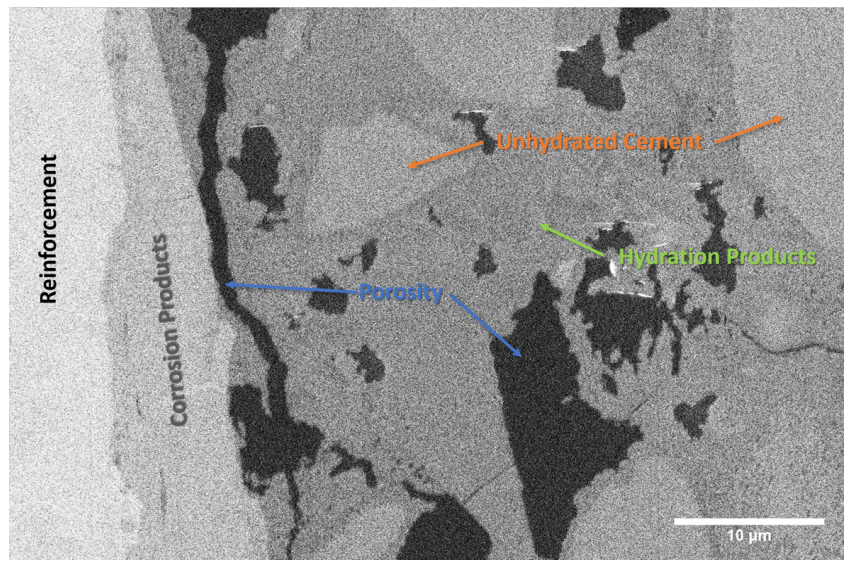

*Figure 7. Tomogram slice (SEM imaged section) of specimen CI. Obtained on 5 UX (ICD, Normal Mode, 4 kV, 3.2 nA, 52° tilt).*

Figure 8 shows an image of a physically polished (see section 2.2) surface of the CI specimen at a random location of the steel-concrete interface obtained by the CBS detector in normal mode at 0° tilt. This image can be compared to the tomogram slice presented in Figure 7, which was a FIB-polished surface. It should be pointed out that the obtained images are similar and similarities in the exhibited microstructures can be observed on both physically and FIB-polished images. The presence of corrosion products at the steel-concrete interface as well as a similar pore size and distribution indicate that the microstructure observed during the tomography is representative of the one that can be commonly observed on the physically polished surface of the specimen.

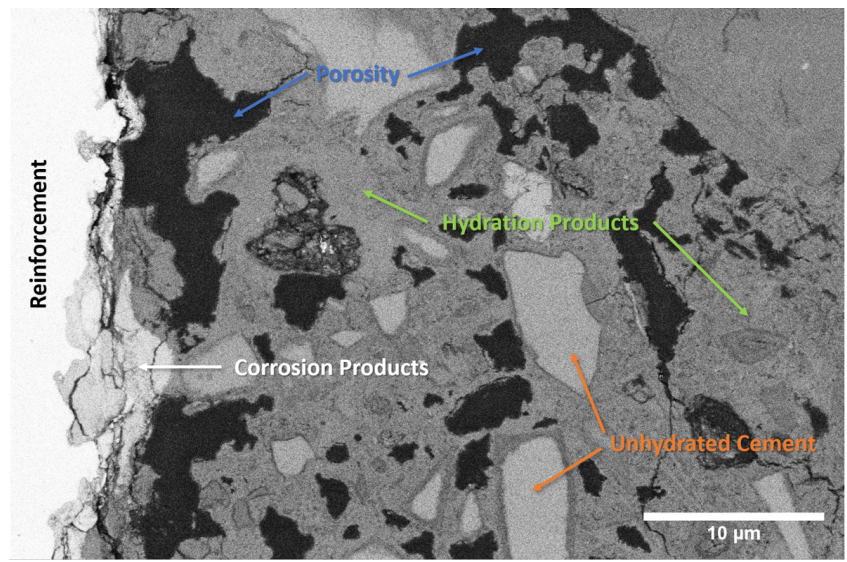

*Figure 8. Surface image of steel-concrete interface of sample CI. Obtained at the sample surface where the tomography trench was opened, taken on 5 UX (CBS, Normal Mode, 4 kV, 3.2 nA, 0° tilt).*

### 3.2. Implications

Figure 7 and Figure 8 show that based on the BSE images that were acquired upon surface preparation by the FIB, and without any further image processing at this stage, different microstructural features can be detected and distinguished, similarly to polished backscattered electron images [10, 39]. The pronounced differences in grey levels that can be observed allow the reliable detection of phases such as voids, cracks, and pore space (black) and steel (white). Furthermore, phases including unhydrated cement particles, hydration products, and corrosion products can be distinguished. This allows for further in-depth analyses of the microstructure, such as by means of algorithms to segment different phases [40] and to quantify characteristics, such as transport properties or volume fractions, size distributions and other stochastic information of features [41]. Since the acquired FIB-SEM tomograms contain stacks of such BSE images and thus provide three-dimensional data, these features can be analysed in 3D, which allows overcoming the limitations of image analysis of sections. Such limitations include, for instance, the difficulty in assessing the size of pores and particles, since a sectional cut rarely is representative of the 3D structure [24]. This shortcoming of information based on 2D sections can be of particular importance when assessing the size of near-spherical features, e.g., entrapped or entrained air voids, where sectional information most likely leads to an underestimation of the actual pore diameter.

The data contained in the tomograms may thus provide a highly valuable basis for future work to obtain detailed insight into 3D features of the interfacial zone. For instance, tomograms of situations representative of "non-corroding interfaces" may be analysed with respect to tortuosity and pore connectivity, or pore diameter distribution. As another example, tomograms of "corroded interfaces" may provide insight into the spatial distribution of corrosion products, both within the cementitious matrix and precipitated within pore space. Such information has previously been collected on two-dimensional BSE maps, sometimes supported by EDX elemental mapping [10]. Since the transport and precipitation of corrosion products is significantly affected by the complex pore structure around the SCI, and distances to which corrosion products can have moved from the steel surface may thus differ from one section to another, we see advantages in obtaining such information in 3D.

The voxel size of the tomograms acquired in this study was in the range 30–50 nm, which is about 100 to 1000 times better than what can be obtained from computed X-ray microtomography [42]. Thus, the FIB-SEM tomograms allow resolving porosity down to the capillary range, particularly to the size scale where capillary suction of water occurs. In terms of studying the durability of reinforced concrete, this is considered highly valuable. It is well known, that capillary water ingress plays a major role in various degradation mechanisms, including freeze-thaw damage, alkali aggregate reactions, and reinforcing steel corrosion [23, 43].

Finally, it should be mentioned that the FIB-SEM tomography technique can be combined with EDX elemental mapping (not done in this study), either on each imaging plane or only on selected ones. Such information would provide valuable complementary data to refine the segmentation of the different phases of interest. For the purpose of creating digital twins of the pore structure, however, such additional information is not considered crucial.

### 3.3. Challenges

While the application of FIB-SEM tomography to specimens at the SCI has the potential to yield valuable experimental data, these experiments also face a few challenges that the authors consider useful to share with the scientific community. Regardless of its rather common utilization in biology and certain fields of material science, its use in studying cementitious materials has been limited [33-35, 44]. Thus, there is limited documented experience about experimental protocols for this specific case.

#### 3.3.1. Tomography volumes

FIB-SEM tomography remains a rather complex, resource and time-consuming process as it requires access to microscopes equipped for this type of analysis for large amount of time [45, 46]. During this work the average slice acquisition time was 3 to 4 minutes, making the tomography pace 1 to 2 hours per micrometer in the third direction depending on the voxel size. As an example, the tomography of CI required approximatively 116 hours of acquisition time spread across 8 microscopy sessions to which about 20 to 25 hours of sample preparation on the microscope were added prior to the tomography itself.

These excessive acquisition times are mainly linked to the rather large volumes that need to be imaged to obtain data that could be considered representative of the sample's microstructure. Microscopes equipped with Gallium Focused Ion Beam are meant for precision tomograms that usually do not exceed a few tens of thousands of cubic micrometers. However, in concrete, the vast range of sizes of features contained in its microstructure require the acquisition of volumes such as CI to obtain a sample's representative volume. Nevertheless, it would seem unreasonable and technically challenging to use this type of equipment to image volumes exceeding 200,000 $\mu m^3$, the acquisition of which would take around 150 h with the currently used techniques. A possible solution for this limitation may be a microscope equipped with a Xenon or multi-species Focused Ion Beam, which would allow a considerably faster milling both during sample preparation and the tomography process.

Another challenge brought by the acquisition of large volumes is the limit that can be reached in terms of image resolution. As mentioned in section 2.3.2, the pixel count of an image is correlated to the horizontal field width (magnification) in order to obtain pixels of a certain size. However, the available imaging pixel amount is technically limited for each microscope which therefore fixes the maximal horizontal field width that can be achieved if the pixel size is fixed accordingly to the desired features' resolution. As an example, if a microscope's maximum image acquisition size is 3,200 by 2,400 pixels, and the desired pixel size is 30 nm for a feature resolution of 90 nm, the theoretical maximum size of the XY plane of the tomogram will be limited to 96 by 72 $\mu m$ respectively. Considering the tomography process software pipeline, and the cropping that usually needs to be done post-acquisition for things such as compensating for an eventual drift, this size will be in the end for exploitable data be reduced by 15 to 30 % in both axes.

### 3.3.2. Immersion mode

As explained in section 2.3.2, the immersion mode allows the acquisition of substantially better-quality images with detectors located in the electron column such as TLD and ICD. However, the use of immersion mode is not possible if the sample contains a large portion of ferro-magnetic material as it uses a localised electro-magnetic field to increase the number of detected electrons (Figure 4). The NCI type specimens contained an austenitic stainless steel, which shows practically no magnetism. Moreover, the steel had the shape of a tube with wall thickness of 250 $\mu m$, which minimized the mass of metal present in the sample. This combination of factors, and primarily the use of non-magnetic austenitic steel, explains why the image acquisition at the SCI was possible in immersion mode in the NCI type of specimens. However, the specimen of type CI contained a carbon steel bar. Since this type of steel is strongly ferromagnetic, the image acquisition in the immersion mode was not possible in this case. Thus, for future studies aiming at characterizing the microstructure of the SCI in a non-corroded state, it may be advisable to use austenitic stainless steels instead of carbon steel, and thus to allow for imaging in the immersion mode.

Several attempts were made to acquire images using the TLD in normal mode, but the image quality was judged insufficient due to the excessive amount of noise and lack of sharpness. This is the reason why the tomogram of the carbon-steel containing sample of type CI was realised on another microscope (5 UX) which allowed the acquisition of acceptable quality images by using the ICD detector in normal mode. Even if this issue is purely equipment related and specific to *FEI/Thermo Fisher* microscopes, it is important to take it into consideration for any eventual future works that would require the usage of FIB-SEM tomography on reinforcement containing concrete.

## 4. Conclusions

In this work, the FIB-SEM nanotomography technique was applied to acquire three-dimensional images (tomograms) of the steel-concrete interfacial zone. In total, five tomograms covering volumes ranging approx. from 8,000 to 200,000 $\mu m^3$ were obtained, whereas four tomograms are representative of non-corroded interfaces, and one of a situation where corrosion products have already precipitated in the interfacial region. The following major conclusions are drawn:

- The obtained tomograms contain stacks of SEM images that allow the observation and characterization of different microstructural features, similarly to polished backscattered electron images. Based on the differences in grey levels, the following phases can be distinguished: steel, unhydrated cement particles, cement hydration products, corrosion products, and pore space including voids and cracks.

- From the stacks of images, three-dimensional images can be generated. This is considered a major advantage as it allows investigating various features in three dimensions, e.g., the pore structure and interconnectivity. In this regard, 2-dimensional images generally available from classical microscopic techniques, looking at sections, tend to yield limited or even misleading information.

- The voxel size of achieved here was in the range 30–50 nm, which is about 100 to 1000 times better than what can be obtained from computed X-ray tomography, especially considering the constraints related to the dimensions of samples that can be reasonably retrieved from the steel-concrete interfacial zone, that is, without affecting and potentially damaging the features of interest during sample preparation.

- For these reasons, FIB-SEM nanotomography allows resolving porosity down to the capillary range as well as distinguishing all features of interest in scientific studies addressing the performance of concrete, particularly the durability and corrosion behaviour of reinforced concrete. The fact that pores at a size scale where capillary suction of water occurs can be resolved is considered a major advantage. Capillary ingress of water is well known to play an important role in various deterioration processes affecting the longevity of reinforced concrete structures.

- The FIB-SEM tomography technique can thus provide datasets as a basis for further analyses of the microstructure, namely by means of algorithms to segment different phases and thus to generate digital twins of the SCI, to further study specific scientific questions, e.g., related to reactive transport or corrosion modelling at the pore scale. For this reason and to maximise the gain of the here acquired data in the scientific community, all tomograms are made freely accessible (see data availability statement below).

- From comparing different techniques for the image acquisition, especially related to the electron detectors, it was here found that the so-called immersion mode can generate substantially better images (in terms of noise) compared to ICD in normal mode. However, the immersion mode cannot be used for specimens that contain ferro-magnetic materials, which thus may often present a problem when studying interfacial regions at (carbon) steel in concrete. Thus, to study non-corroded situations, we recommend the use of non-magnetic stainless steel as a model system. Alternatively, if carbon steel is used, the removal of noise arising from the ICD should receive special attention in later image processing.

## Acknowledgements

The authors are grateful to the European Research Council (ERC) for the financial support provided under the European Union's Horizon 2020 research and innovation program (grant agreement no. 848794) and to the Swiss National Science Foundation for the financial support (project no. 196919). The authors gratefully acknowledge ScopeM, ETH Zurich's scientific center for optical and electron microscopy, for their support & assistance in this work. The authors wish to thank Dr. Michele Griffa (EMPA, Switzerland) for the valuable discussions and his input.

## Author contributions

U.A. and N.R. conceived the study. Experimental work was performed by N.R.; Z.Z. provided the reinforced concrete specimens. N.R., U.A. and T.S. performed the analysis and interpretation of the results. N.R. and U.A. wrote the main draft of the manuscript, to which all authors contributed. U.A. was the main supervisor of the project. All authors read and approved the final manuscript.

## Declaration of interests

The authors declare that they have no known competing financial interests or personal relationships that could have appeared to influence the work reported in this paper.

## Data availability

All the tomograms acquired in this study are available for open access download: doi.org/10.5281/zenodo.8192942.

# References

[1] U.M. Angst, M.R. Geiker, A. Michel, C. Gehlen, H. Wong, O.B. Isgor, B. Elsener, C.M. Hansson, R. Francois, K. Hornbostel, R. Polder, M.C. Alonso, M. Sanchez, M.J. Correia, M. Criado, A. Sagues, N. Buenfeld, The steel-concrete interface, Mater Struct, 50 (2017) 143.

[2] U.M. Angst, M.R. Geiker, M.C. Alonso, R. Polder, B. Elsener, O.B. Isgor, H. Wong, A. Michel, K. Hornbostel, C. Gehlen, R. François, M. Sanchez, M. Criado, H. Sørensen, C. Hansson, R. Pillai, S. Mundra, J. Gulikers, M. Raupach, J. Pacheco, A. Sagüés, The effect of the steel-concrete interface on chloride-induced corrosion initiation in concrete – a critical review by RILEM TC 262-SCI, Mater Struct, 52 (2019) 88.

[3] C.L. Page, Initiation of chloride-induced corrosion of steel in concrete: role of the interfacial zone, Mater Corros, 60 (2009) 586-592.

[4] D.A. Hausmann, Steel corrosion in concrete. How does it occur?, Materials Protection, 6 (1967) 19-23.

[5] V.K. Gouda, Corrosion and corrosion inhibition of reinforcing steel. I. Immersed in alkaline solutions., Br. Corros. J., 5 (1970) 198-203.

[6] C.L. Page, Mechanism of corrosion protection in reinforced concrete marine structures, Nature, 258 (1975) 514-515.

[7] C.L. Page, Ø. Vennesland, Pore solution composition and chloride binding capacity of silica fume-cement pastes, Mater Struct, 19 (1983) 19-25.

[8] B. Huet, V. L'Hostis, L. Tricheux, H. Idrissi, Influence of alkali, silicate, and sulfate content of carbonated concrete pore solution on mild steel corrosion behavior, Materials and Corrosion-Werkstoffe Und Korrosion, 61 (2010) 111-124.

[9] M. Stefanoni, U. Angst, B. Elsener, Influence of Calcium Nitrate and Sodium Hydroxide on Carbonation-Induced Steel Corrosion in Concrete, Corros, 75 (2019) 737-744.

[10] H.S. Wong, Y.X. Zhao, A.R. Karimi, N.R. Buenfeld, W.L. Jin, On the penetration of corrosion products from reinforcing steel into concrete due to chloride-induced corrosion, Corrosion Science, 52 (2010) 2469-2480.

[11] M. Stefanoni, U.M. Angst, B. Elsener, Kinetics of electrochemical dissolution of metals in porous media, Nat Mater, 18 (2019) 942-+.

[12] B. Reddy, Influence of the steel-concrete interface on the chloride threshold level, Imperial College, London, 2001.

[13] A. Castel, T. Vidal, R. François, G. Arliguie, Influence of steel-concrete interface quality on reinforcement corrosion induced by chlorides, Mag. Concr. Res., 55 (2003) 151-159.

[14] T.A. Soylev, R. François, Quality of steel-concrete interface and corrosion of reinforcing steel, Cement and Concrete Research, 33 (2003) 1407-1415.

[15] J.G. Nam, W.H. Hartt, K. Kim, Effects of air void at the steel-concrete interface on the corrosion Initiation of reinforcing steel in concrete under chloride exposure, Journal of the Korea Concrete Institute, 17 (2005) 829-834.

[16] R. François, I. Khan, N.A. Vu, H. Mercado, A. Castel, Study of the impact of localised cracks on the corrosion mechanism, Eur. J. Environ. Civ. Eng., 16 (2012) 392-401.

[17] E. Rossi, R. Polder, O. Copuroglu, T. Nijland, B. Savija, The influence of defects at the steel/concrete interface for chloride-induced pitting corrosion of naturally-deteriorated 20-years-old specimens studied through X-ray Computed Tomography, Construction and Building Materials, 235 (2020).

[18] S. Robuschi, A. Tengattini, J. Dijkstra, I. Fernandez, K. Lundgren, A closer look at corrosion of steel reinforcement bars in concrete using 3D neutron and X-ray computed tomography, Cement and Concrete Research, 144 (2021).

[19] J.A. Gonzalez, S. Algaba, C. Andrade, Corrosion of Reinforcing Bars in Carbonated Concrete, Br. Corros. J., 15 (1980) 135-139.

[20] C. Alonso, C. Andrade, J.A. González, Relation between resistivity and corrosion rate of reinforcements in carbonated mortar made with several cement types, Cement and Concrete Research, 8 (1988) 687-698.

[21] C. Andrade, A. Castillo, Evolution of reinforcement corrosion due to climatic variations, Mater Corros, 54 (2003) 379-386.

[22] M. Stefanoni, U.M. Angst, B. Elsener, Electrochemistry and capillary condensation theory reveal the mechanism of corrosion in dense porous media, Scientific Reports, 8 (2018) 7407.

[23] Z. Zhang, P. Trtik, F. Ren, T. Schmid, C.H. Dreimol, U. Angst, Dynamic effect of water penetration on steel corrosion in carbonated mortar: A neutron imaging, electrochemical, and modeling study, CEMENT, 9 (2022) 100043.

[24] H.S. Wong, U.M. Angst, M.R. Geiker, O.B. Isgor, B. Elsener, A. Michel, M.C. Alonso, M.J. Correia, J. Pacheco, J. Gulikers, Y.X. Zhao, M. Criado, M. Raupach, H. Sorensen, R. Francois, S. Mundra, M. Rasol, R. Polder, Methods for characterising the steel-concrete interface to enhance understanding of reinforcement corrosion: a critical review by RILEM TC 262-SCI, Mater Struct, 55 (2022).

[25] A. Cesen, T. Kosec, A. Legat, Characterization of steel corrosion in mortar by various electrochemical and physical techniques, Corrosion Science, 75 (2013) 47-57.

[26] M. Beck, J. Goebbels, A. Burkert, Application of X‐ray tomography for the verification of corrosion processes in chloride contaminated mortar, Mater Corros, 58 (2007) 207-210.

[27] G. Ebell, A. Burkert, J. Fischer, J. Lehmann, T. Muller, D. Meinel, O. Paetsch, Investigation of chloride-induced pitting corrosion of steel in concrete with innovative methods, Materials and Corrosion-Werkstoffe Und Korrosion, 67 (2016) 583-590.

[28] P. Zhang, Z.L. Liu, Y. Wang, J.B. Yang, S.B. Han, T.J. Zhao, 3D neutron tomography of steel reinforcement corrosion in cement-based composites, Construction and Building Materials, 162 (2018) 561-565.

[29] Z. Zhang, M. Shakoorioskooie, M. Griffa, P. Lura, U.M. Angst, A laboratory investigation of cutting damage to the steel-concrete interface, Cement and Concrete Research, 138 (2020) 106229.

[30] G.R. Davis, J.C. Elliott, Artefacts in X-ray microtomography of materials, Mater Sci Techn, 22 (2006) 1011-1018.

[31] B. Šavija, M. Luković, S.A.S. Hosseini, J. Pacheco, E. Schlangen, Corrosion induced cover cracking studied by X-ray computed tomography, nanoindentation, and energy dispersive X-ray spectrometry (EDS), Mater Struct, 48 (2015) 2043-2062.

[32] A. Mouton, N. Megherbi, K. Van Slambrouck, J. Nuyts, T.P. Breckon, An experimental survey of metal artefact reduction in computed tomography, Journal of X-Ray Science and Technology, 21 (2013) 193-226.

[33] L. Holzer, Quantification of capillary porosity in cement paste using high resolution 3D-microscopy: potential and limitations of FIB-nanotomography, in: J. Marchand (Ed.) 2nd Int. RILEM Symp. on Advances in Concrete through Science and EngineeringQuebec City, Canada, 2006.

[34] L. Holzer, B. Muench, M. Wegmann, P. Gasser, R.J. Flatt, FIB-Nanotomography of particulate systems - Part I: Particle shape and topology of interfaces, J. Am. Ceram. Soc., 89 (2006) 2577-2585.

[35] B. Munch, P. Gasser, L. Holzer, R. Flatt, FIB-nanotomography of particulate systems - Part II: Particle recognition and effect of boundary truncation, J. Am. Ceram. Soc., 89 (2006) 2586-2595.

[36] J.A. Taillon, C. Pellegrinelli, Y.L. Huang, E.D. Wachsman, L.G. Salamanca-Riba, Improving microstructural quantification in FIB/SEM nanotomography, Ultramicroscopy, 184 (2018) 24-38.

[37] H. Nyquist, Certain topics in telegraph transmission theory (Reprinted from Transactions of the A. I. E. E., February, pg 617-644, 1928), P Ieee, 90 (2002) 280-305.

[38] J.B. Pawley, Chapter 4. Points, Pixels, and Gray Levels: Digitizing Image Data, Handbook of Biological Confocal Microscopy, Springer, 2006, pp. 59-79.

[39] K.L. Scrivener, Backscattered electron imaging of cementitious microstructures: understanding and quantification, Cement & Concrete Composites, 26 (2004) 935-945.

[40] H.S. Wong, M.K. Head, N.R. Buenfeld, Pore segmentation of cement-based materials from backscattered electron images, Cement and Concrete Research, 36 (2006) 1083-1090.

[41] H.S. Wong, N.R. Buenfeld, M.K. Head, Estimating transport properties of mortars using image analysis on backscattered electron images, Cement and Concrete Research, 36 (2006) 1556-1566.

[42] S. Brisard, M. Serdar, P.J.M. Monteiro, Multiscale X-ray tomography of cementitious materials: A review, Cement and Concrete Research, 128 (2020).

[43] L. Bertolini, B. Elsener, P. Pedeferri, E. Redaelli, R. Polder, Corrosion of Steel in Concrete: Prevention, Diagnosis, Repair (2nd Edition), WILEY VCH, Weinheim, 2013.

[44] N. Ruffray, From fresh concrete microstructure to digitally fabricated HPFRC: A challenging journey up from the nanoscale in search of precious digital macro-applications, ETH Diss. No. 26937, ETH Zurich, Switzerland, 2020.

[45] C.J. Peddie, C. Genoud, A. Kreshuk, K. Meechan, K.D. Micheva, K. Narayan, C. Pape, R.G. Parton, N.L. Schieber, Y. Schwab, B. Titze, P. Verkade, A. Weigel, L.M. Collinson, Volume electron microscopy, Nature Reviews Methods Primers, 2 (2022) 51.

[46] L.M. Collinson, C. Bosch, A. Bullen, J.J. Burden, R. Carzaniga, C. Cheng, M.C. Darrow, G. Fletcher, E. Johnson, K. Narayan, C.J. Peddie, M. Winn, C. Wood, A. Patwardhan, G.J. Kleywegt, P. Verkade, Volume EM: a quiet revolution takes shape, Nature Methods, 20 (2023) 777-782.